\documentclass[aps,english,twocolumn]{revtex4}
\usepackage{mathrsfs}
\usepackage{graphicx}
\usepackage{amsmath, amssymb}
\usepackage{babel}

\begin{document}

 \title{ In-gap bound states and tunnelling conductance of multi-band superconductors through a $N-S_1-S_2$ junction}
 \author{Xiao-Yong Feng and Tai-Kai Ng}
\affiliation{Department of Physics, Hong Kong University of Science
and Technology, Clear Water Bay Road, Hong Kong}

 \begin{abstract}
  The tunnelling conductance between a metal and a multi-band $s$-wave superconductor with a thin layer of single-band
 $s$-wave superconductor sandwiched in between is examined in this paper. We show that a in-gap peak in conductance curve
 is found as a result of the formation of in-gap bound state between the single-band and multi-band superconductors
 junction if the phases of the superconducting order parameters of the multi-band superconductor is frustrated. The
 implication of this result in determining the gap symmetry of the iron-based superconductors is discussed.
 \end{abstract}

\maketitle

  With the discovery of the Iron-based (pnictides) superconductors, the superconductivity characterized by more than one
 order parameters, i.e. the multi-gap superconductors, becomes a hot topic. Band structure calculations indicate that the
 material has a quasi-two-dimensional electronic structure, with five bands centered around the $\Gamma$- and $M$- points
 in the Brillouin zone contributing to the Fermi surface.  An immediate issue of interests is thus the symmetry of the
 superconductor order parameters. Besides the obvious possibility that all the bands share the same superconductor
 order parameter symmetry, it is also possible that the order parameters pick up more complicated configurations because
 of the non-trivial Fermi surface structure. For example, it has been proposed that the order parameters may have
 $s$-wave symmetry, but with opposite sign between bands centered at $\Gamma$- and $M$-points\cite{hu,mazin,wang}. Other more
 exotic possibilities have also been proposed\cite{kuroki}.

  Experimentally, the absent of $\pi$ phase shift between tunnelling in different direction in scanning SQUID
 microscopy\cite{SQUID} and the suppression of knight shift with decreasing temperature in NMR\cite{NMR1} experiments
 seems to exclude spin-triplet pairing states. However controversy remains between the $d$-wave and the exotic
 $s$-wave symmetries. The d-wave symmetry gets support from NMR\cite{NMR1} and lower critical field
 data\cite{field} indicating existence of nodes in the superconducting order parameter while the observations in
 Angle-resolved photoelectron spectroscopy\cite{ARPES1,ARPES2} favor node-less gaps.

\begin{figure}[h]
 \includegraphics [width=4cm]{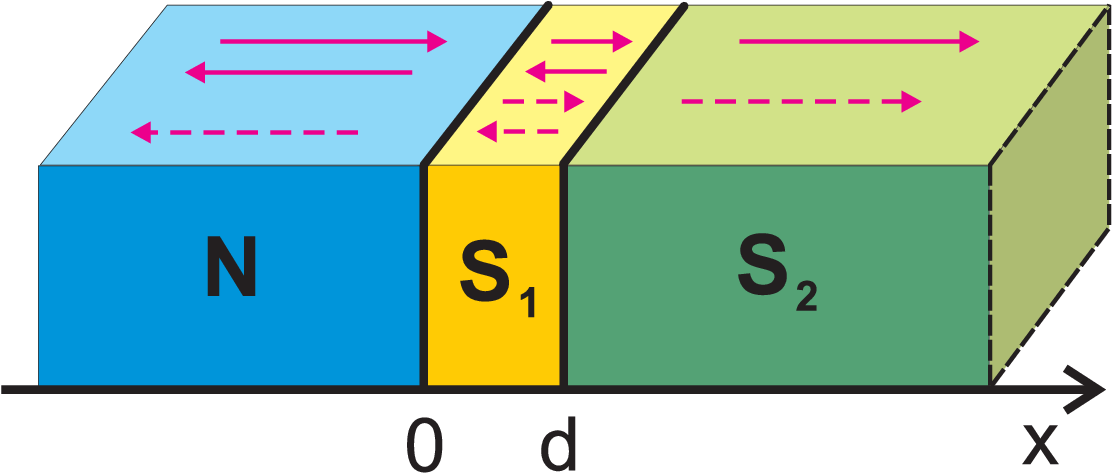}
 \caption{The geometry of the $N-S_1-S_2$ junction. The solid (dashed) arrows
 denote the propagation direction of particles (holes). }
 \label{geometry}
 \end{figure}

  Thus an experiment which can distinguish the different $d$-wave and $s$-wave gap symmetry possibilities is significant. In
  this paper, we show that the quasi-particle tunnelling through a $N-S_{1}-S_{2}$ junction where $N$, $S_{1}$ and
  $S_{2}$ represents normal metal, a single-band $s$-wave superconductor and the targeted superconductor respectively,
  provide a strong test to the above problem when tunnelling data from different surfaces of target superconductor $S_2$
  and with different thickness $S_1$ are collected. The geometry of the junction is sketched in figure \ref{geometry}.

  To begin with, we first review the situation of simple $s$-wave and $d$-wave superconductors. We shall assume that
  the superconductor order parameter structure is prominent on the $x-y$ plane, and superconductivity along
  $z$-direction mainly comes from Josephson effect, consistent with a quasi-two-dimensional band structure. For the
  $s$-wave superconductor, it is expected that the tunnelling spectrum will be qualitatively the same independent of
  the tunnelling plane of the targeted superconductor and the thickness ($d$) of $S_1$. The situation is however, very
  different for the $d$-wave superconductor. The existence of a mid-gap state at the boundary between the $d$-wave
  superconductor and normal metal if the surface is cut along the $(110)$ direction\cite{midgap} leads to a zero-bias
  peak in the tunnelling spectrum along the $(110)$ direction. Thus the existence of a zero-bias peak in tunnelling
  experiment with one prominent surface orientation distinguishes between conventional $s$- and $d$-wave
  superconductors\cite{Tanaka}.

   Next we consider a multi-band $s$-wave superconductor with order parameters of opposite sign. We shall call it $\pm
  s$-wave superconductor in the following. It was pointed out in Ref.\cite{Ng}, that in-gap bound states exist at the
  interface between a single-band $s$-wave superconductor and a $\pm s$-wave superconductor and it is expected that the
  in-gap bound states will contribute to quasi-particle tunnelling leading to in-gap peaks in the tunnelling conductance
  through an $N-S_{1}-S_{2}$ junction. We shall show in the following that the $\pm s$-wave superconductor has a
  rather unique tunnelling spectrum that distinguishes itself from ordinary
  $s$- or $d$-wave superconductors.

   The quasi-particle behavior of the multi-band system can be obtained by solving the Bogoliubov-de Gennes (BdG)
   equations\cite{BdG}
 \begin{equation}
 \label{bdg}
 \epsilon^{(i)}\left(
 \begin{array}{c}
   u^{(i)}(\vec{r}) \\ v^{(i)}(\vec{r})
 \end{array}
 \right)=\left(
 \begin{array}{cc}
  \hat{H}^{(i)}-\mu & \Delta^{(i)}(\vec{r}) \\
  \Delta^{(i)*}(\vec{r}) & -\hat{H}^{(i)}+\mu
  \end{array}\right)\left(
  \begin{array}{c}
  u^{(i)}(\vec{r}) \\ v^{(i)}(\vec{r})
 \end{array}\right)
 \end{equation}
  where $i=1,..N$, $N$ is the number of contributing bands in the multi-band superconductor, $\hat{H}^{(i)}$ is the
  band dispersion of band $i$ and $\Delta^{(i)}(\vec{r})=|\Delta^{(i)}(x)|e^{i\theta^{(i)}(x)}$ is the corresponding
  superconductor order parameter. $\mu$ is the chemical potential. We have assumed that quasi-particles at different bands
  are decoupled from each other in writing down Eq.\ (\ref{bdg}).

   To model the $N-S_1-S_2$ junction, we assume $(|\Delta^{(i)}(x,y,z)|,\theta^{(i)}(x,y,z))$ = $(0,0)$, $(\Delta_{1},0)$
  and $(\Delta_{2}^{(i)},\theta^{(i)})$ in the regions $x<0$, $0<x<d$ and $x>d$, respectively. The different regions are
  connected by $\delta$-function scattering potentials $U_1\delta(x)$ and $U_2\delta(x-d)$, respectively and the system
  is assumed to be translational invariant in $(y,z)$ direction.  We also assume that the $\delta$-function scattering
  potentials do not introduce scattering between different bands, so that Eq.\ (\ref{bdg}) remains valid in the presence
  of the junction. Within this approximation the total tunnelling current is just the
  sum of tunnelling currents from all bands. Notice that we are allowed to choose the phase of the single-band
  superconductor $S_1$ to be $\theta(0<x<d)=0$ since overall phase of the system is a pure gauge. On the other hand there
  are in general $N$ different phases $\theta^{(i)}(x>0)$ in the multi-band superconductor $S_2$. We shall assume that
  only two groups of $\theta^{(i)}$'s exist, $\theta^{(i)}=(\theta^A,\theta^B)$, with $\theta^A=\theta^B+\pi$ in the bulk
  multi-band superconductor ($\pm s$-wave), as is proposed to be the case in the pnictides superconductors.

   The Josephson coupling between a single-band superconductor and a $\pm s$-wave superconductor has been analyzed in
  Ref.\cite{Ng}. Let $T_{A(B)}$ be the Josephson coupling of the first(second) group of bands to $S_1$. It was found
  that when $T_A>>(<<)T_B$, then $\theta_{A(B)}=0$ and $\theta_{B(A)}=\pi$. This state respect time-reversal symmetry
  and is called the TRI state. An alternative solution exist when $T_A\sim T_B$, at which case $\theta_A,\theta_B\neq0,
  \pi$. The state breaks time-reversal symmetry and is called the TRB state. We first consider the tunnelling
  conductance between the $s$-wave superconductor and band $i$ in the $\pm s$-wave
  superconductor with $\theta^{(i)}=\theta$.

   Eq.\ (\ref{bdg}) have plane-wave solutions of form
 \begin{eqnarray}
 \label{ssol}
 \Psi(\vec{r})=\left(\begin{array}{c}
           u(\vec{r})\\
           v(\vec{r})
         \end{array}\right)=e^{i\vec{q}_{\|}\cdot\vec{r}_{\|}}e^{iqx}\left(\begin{array}{c}
           u_{q}\\
           v_{q}
         \end{array}\right)
 \end{eqnarray}
  where we have neglected the band index $i$ for brevity. $u_{q}$ and $v_{q}$ satisfy
 \begin{equation}
 \epsilon\left(\begin{array}{c}
           u_q\\
           v_q
         \end{array}\right)=\left(\begin{array}{cc}
         \xi_{\vec{q}} & \Delta \\
         \Delta & -\xi_{\vec{q}}
         \end{array}\right)
         \left(\begin{array}{c}
           u_{q}\\
           v_{q}
         \end{array}\right)
 \end{equation}
 where $\xi_{\vec{q}}=\varepsilon_{\vec{q}}-\mu$ and $\varepsilon_{\vec{q}}$ is the band dispersion. We shall assume
 $\varepsilon_{\vec{q}}=\hbar^2(q^2+q_{\|}^2)/2m$ for all bands and $\mu=\hbar^2k_f^2/2m$ in our calculation where
 $k_f$ is the Fermi momentum and shall consider the semi-classical limit $k_f,m\rightarrow\infty$ with $v_f=\hbar k_f/m$
 remaining finite such that band structure effects are eliminated in our calculation.

  For a given energy $\epsilon$ and transverse momentum $q_{\|}$, there are four possible longitudinal momenta
  $q=\pm q_+,\pm q_-$, where $\hbar q_{+(-)}=\sqrt{2m\left(\mu_{x}+(-)\sqrt{\epsilon^{2}-|\Delta|^{2}}\right)}$ and
  $\mu_{x}=\mu-\hbar^2q_{\|}^2/2m$. The corresponding $u_q,v_q$ is given by
  \begin{eqnarray}
  \left(\begin{array}{c}
           u_{q}\\
           v_{q}
         \end{array}\right)=\left(\begin{array}{c}
           \sqrt{\frac{1}{2}(1+\frac{\xi_{\vec{q}}}{\epsilon})}\\
            e^{-i\theta}\sqrt{\frac{1}{2}(1-\frac{\xi_{\vec{q}}}{\epsilon})}
                    \end{array}\right).
  \end{eqnarray}

 Applying these results to our junction problem the general scattering wave-function of an incoming wave from metal
 side has the following form at various regions in space,
 \begin{eqnarray}
 \label{solf}
 \Psi_{N}(x)e^{-i\vec{q}_{\|}\cdot\vec{r}_{\|}}&=&(e^{ik_{+}x}+a_{1}e^{-ik_{+}x})\left(\begin{array}{c}
           1 \\
           0
         \end{array}\right) +a_{2}e^{ik_{-}x}\left(\begin{array}{c}
           0 \\
           1
         \end{array}\right)  \nonumber \\
 \Psi_{S_{1}}(x)e^{-i\vec{q}_{\|}\cdot\vec{r}_{\|}}&=&(a_{3}e^{ip_{+}x}+a_{4}e^{-ip_{+}x})
 \left(\begin{array}{c}
           u_{p_{+}}\\
           v_{p_{+}}
         \end{array}\right)\nonumber\\
         &&+(a_{5}e^{-ip_{-}x}+a_{6}e^{ip_{-}x}) \left(\begin{array}{c}
           u_{p_{-}}\\
           v_{p_{-}}
         \end{array}\right) \nonumber \\
 \Psi_{S_{2}}(x)e^{-i\vec{q}_{\|}\cdot\vec{r}_{\|}}&=&a_{7}e^{iq_{+}x} \left(\begin{array}{c}
           u_{q_{+}}\\
           v_{q_{+}}
         \end{array}\right)
         +a_{8}e^{-iq_{-}x} \left(\begin{array}{c}
           u_{q_{-}}\\
           v_{q_{-}}
         \end{array}\right)
 \end{eqnarray}
 where
 \begin{eqnarray}
 \hbar k_{+(-)}&=& \sqrt{2m\left(\mu_{x}+(-)|\epsilon|\right)}  \\ \nonumber
 \hbar p_{+(-)}&=& \sqrt{2m\left(\mu_{x}+(-)\sqrt{\epsilon^{2}-|\Delta_1|^2}\right)} \\ \nonumber
 \hbar q_{+(-)}&=& \sqrt{2m\left(\mu_{x}+(-)\sqrt{\epsilon^{2}-|\Delta_2|^2}\right)}.
 \end{eqnarray}
 Notice that $\mu_x$ has the same value at all regions because of momentum conservation along $y-z$ directions.
 However $\Delta_2\rightarrow\Delta_2^{(i)}$ and $q_{+(-)}^{(i)}$'s are in general different for different bands.

 The wavefunction has to satisfy the boundary conditions
 \begin{eqnarray}
 \Psi_{N}(0)-\Psi_{S_{1}}(0)&=&0 \\  \nonumber
 \Psi_{N}'(0)-\Psi_{S_{1}}'(0)&=&-\frac{2mU_{1}}{\hbar^2}\Psi_{N}(0) \\ \nonumber
 \Psi_{S_{1}}(d)-\Psi_{S_{2}}(d)&=&0 \\ \nonumber
 \Psi_{S_{1}}'(d)-\Psi_{S_{2}}'(d)&=&-\frac{2mU_{2}}{\hbar^2}\Psi_{S_{1}}(d)
 \end{eqnarray}
 at the interfaces which determine the eight coefficients $\{a_{n}\}$ entering Eq.\ (\ref{solf}). The tunnelling
 conductance defined by $dI/dV$ is proportional to $\int dq_{\|}\left(k_+(1-|a_{1}(\epsilon=eV)|^{2})+k_-|a_{2}
 (\epsilon=eV)|^{2}\right)$. In the semi-classical limit $k_f>>q_{\|}$, the main contribution to the tunnelling current
 comes from electrons with normal incidence to the interface.
 In this case we may approximate
 $k_+\sim k_-\sim k_f, \mu_x\sim \mu$ and the BTK result
 \begin{equation}
 \label{btk}
 {dI\over dV}\sim1-|a_{1}(\epsilon=eV)|^{2}+|a_{2}(\epsilon=eV)|^{2}
 \end{equation}
  is recovered\cite{BTK}. we shall employ this approximation in the following analysis.

\begin{figure}[h]
 \includegraphics [width=6cm]{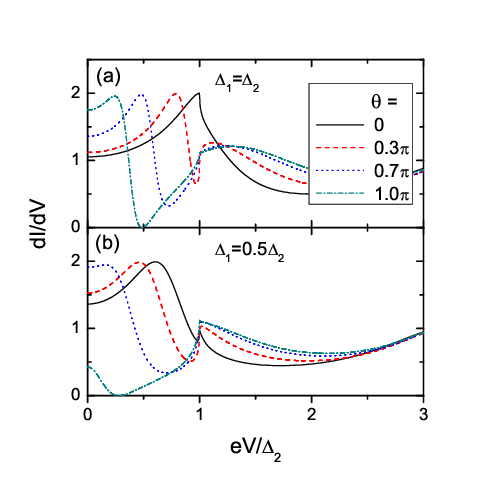}
 \caption{Tunnelling conductance for various $\theta$. (a)For
 $\Delta_{1}=\Delta_{2}=0.001\mu$; (b)For $\Delta_{1}=0.5\Delta_{2}=0.0005\mu$. $U_{1}=U_{2}=\mu$ and $d=\xi_{c2}$
 in both cases,. } \label{theta}
 \end{figure}

   To see the existence of bound states in the $S_1-S_2$ interface we first look at the $\theta$-dependence of the tunnelling
   conductance. In figure\ref{theta} we plot $dI/dV$ versus $V$ for various values of $\theta$. we have chosen
  $\Delta_2=0.001\mu$, $U_1=U_2=\mu$ and $d=\xi_{c2}$, where $\xi_{c2}\sim v_f/\Delta_2$ is the coherence length
  corresponding to superconducting gap $\Delta_2$ in generating the figure. First we consider $\theta=0$ which represents
  tunnelling between usual (un-frustrated) $s$-wave superconductor junctions. In this case the conductance is peaked at
  the gap edges $\sim\Delta_1,\Delta_2$ and there is no in-gap peak. The peak at the lower gap edge $\Delta_1$ moves
  towards zero energy as $\theta$ increases, corresponding to the appearance of TRB states. Notice there is no conductance peak
  between $\Delta_1$ and $\Delta_2$.

  Notice that instead of approaching zero energy, the conductance peak saturates at a finite value $eV>0$ when
  $\theta\rightarrow\pi$. This is because we have employed a finite thickness $d\sim\xi_{c2}$ in our calculation. In this
  case the incoming and reflected waves inside $S_1$ interfere and split the zero-energy bound states into two states with
  energy $\epsilon\sim\pm\Delta_1 e^{-\xi_{c1}/d}$.
 \begin{figure}[h]
 \includegraphics [width=7cm]{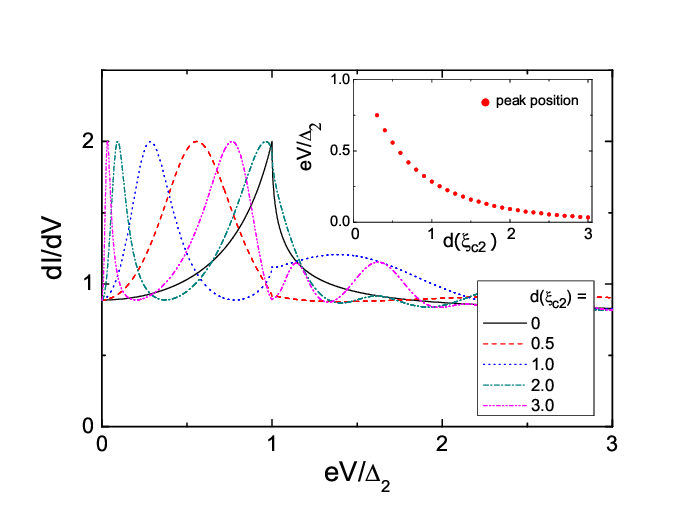}
 \caption{Tunnelling conductance for various $d$.
 $\Delta_{1}=\Delta_{2}=0.001\mu$, $U_{1}=\mu$ and $U_{2}=0$. The inset
 is d-dependent of the conductance peak position. } \label{d}
 \end{figure}

  The thickness dependence of conductance is examined in figure \ref{d} where we have computed the tunnelling spectrum for
 $\Delta_1=\Delta_2$ and $\theta=\pi$ from $d=0$ to $d\sim3\xi_{c2}$. When $d=0$, the tunnelling behavior is the same as
 the case of normal metal tunnelling to a s-wave superconductor where no in-gap bound state exists. As $d$ increases, we
 see that an in-gap peak in conductance is developed with peak position moving towards zero bias. The peak also becomes
 sharper as $d$ increases. The position of the peak as function of $d$ is shown in the inset of figure \ref{d} where it is
 clear that the peak position approaches zero as $d\rightarrow\infty$\cite{Ng}. Oscillatory behavior is also found in the
 conductance curves at energy $eV>\Delta_{1}$ at large $d$. This is a result of interference between multiple reflecting
 waves and is observable only if the total path length ($\sim$ a few $d$) is smaller than the inelastic scattering
 mean-free path ($l$) of the quasi-particles. We note the existence of the in-gap conductance peak is independent of
 the surface orientation for an $\pm s$-wave superconductor. This and the $d$ dependence of the in-gap conductance peak
 provides a clear distinction between an ordinary $d$-wave and $\pm s$-wave superconductors.

  \begin{figure}[h]
  \includegraphics [width=6cm]{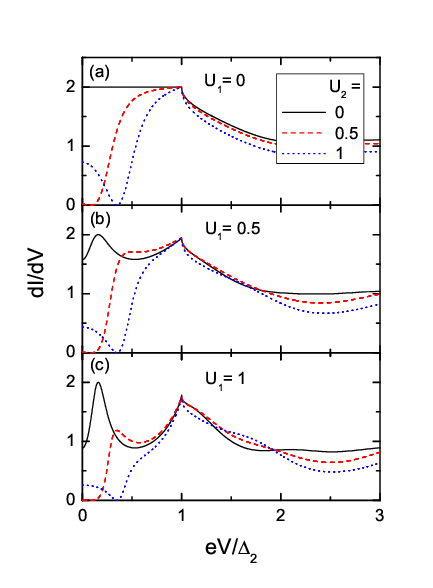}
  \caption{The conductance with various contact barriers. The other
  parameters are taken as $\Delta_{2}=2\Delta_{1}=0.001\mu$ and $d=1.5\xi_{c2}$.
  } \label{Z}
  \end{figure}

   The quantitative behavior of the tunnelling conductance depends also strongly on the scattering parameters $U_1$ and
 $U_2$. Figure \ref{Z} shows the $dI/dV-V$  curves with various contact barrier $U_{1}$ and $U_{2}$ for $\theta=\pi$ and
 $\Delta_1=0.5\Delta_2$. First we note that with $U_{1}$ fixed, the conductance is generally suppressed by $U_{2}$ at
 energy $eV>\Delta_1$. At this regime the quasi-particles do not see superconductor $S_1$ and tunnelling spectrum
 is qualitatively the same as the one from a normal metal directly to a $s$-wave superconductor. We also observe that the
 conductance peak within $\Delta_{1}$ originating from the in-gap bound state exists only for nonzero $U_{1}$ and becomes
 more pronounced as $U_1$ increases. This is because the incoming waves couple to the in-gap bound state only when
 $U_1$ is nonzero.

 \begin{figure}[h]
  \includegraphics [width=6cm]{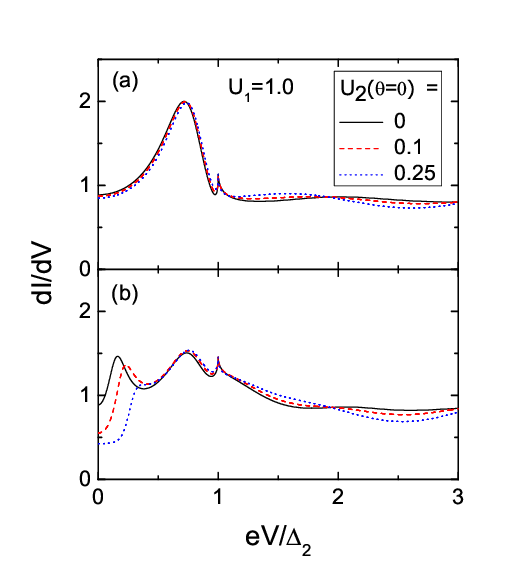}
  \caption{The total conductance when (a)$S_{2}$ is a $s$-wave superconductor; (b) $S_{2}$ is a $\pm s$-wave
  superconductor. $\Delta_{1}=0.5\Delta_{2}=0.0005\mu$, $U_{1}=\mu$ and $d=1.5\xi_{c2}$ in both cases.
  $U_{2}(\theta^{(i)}=\pi)=2U_{2}(\theta^{(i)}=0)$ in the case of $\pm s$-wave superconductor.}
  \label{add3}
  \end{figure}

  We shall now examine the TRI state in more detail. In this case there are two groups of tunnelling channels
  $-$ one with $\theta^{(i)}=0$ and the other with $\theta^{(i)}=\pi$. $S_{1}$ is strongly coupled with the
  $\theta^{(i)}=0$ bands (smaller $U_2$) and has weaker coupling (large $U_2$) to the $\theta^{(i)}=\pi$ bands. The
  tunnelling conductance is a sum of contributions from all bands. To compute the tunnelling spectrum we shall assume
  that the two groups of bands have the same gap magnitude $\Delta_2$, and the $s$-wave superconductor has a weaker gap
  $\Delta_1=0.5\Delta_2$. We shall also assume that $U_2(\theta^{(i)}=\pi)=2U_{2}(\theta^{(i)}=0)$.  The two bands
  contribute equally to the conductance in our calculation. As shown in figure \ref{add3}, conductance peak
  below $\Delta_{1}$ appears only when $S_{2}$ is a $\pm s$-wave superconductor. The conductance of the $s$-wave
  superconductor is rather insensitive to $U_{2}$ while the in-gap conductance peak of the $\pm s$-wave superconductor is
  sensitive to changing $U_{2}$. In particular, smaller $U_{2}$ favors a sharper peak. This is another unique feature
  distinguishing the $\pm s$-wave superconductivity from other symmetries.

  Experimentally the tunnelling spectrum of a multi-band superconductor can be obtained rather straightforwardly from STM
 experiments. The $s$-wave superconductor can be introduced by coating the STM tip with a thin layer of the $s$-wave
 superconductor and our predictions can be tested by performing the experiment with different thickness of coated $s$-wave
 superconductor on different surfaces of the targeted superconductor. The tunnelling barrier $U_2$ can be tuned by
 varying the distance between the STM tip and the targeted superconductor.
 We note however that nature is much more complicated than what our simple calculation presents which ignores the
 complicated band structures of the material, surface details and scattering between different bands. Nevertheless the
 strong distinctive features in the tunnelling spectrum of $\pm s$-wave superconductor we predict here should be
 detectable in experiment\cite{jhu} and provides a strong test to the gap symmetry of the Iron-base superconductors.

  We thank Prof. H.H. Wen for interesting discussions during the course of this work. This work is supported by HKUGC
  through grant CA05/06.Sc04.

\end{document}